\documentclass{appolb}
\usepackage{graphicx}
\usepackage{caption}
\usepackage{subcaption}
\begin{document}

\title{Results from STAR Beam Energy Scan Program 
\thanks{Presented at 42. International Symposium on Multiparticle Dynamics, Kielce, Poland, 16-21 September 2012.}}%
\author{Michal \v{S}umbera (for the STAR Collaboration)
\address{Nuclear Physics Institute ASCR, 250 68 \v{R}e\v{z}, Czech Republic}}
\maketitle
\begin{abstract}
Results from the Beam Energy Scan (BES) program conducted recently by STAR experiment at RHIC are presented.  The data from Phase-I of the BES program collected in  Au+Au collisions at center-of-mass energies ($\sqrt{s_{NN}}$) of  7.7, 11.5, 19.6, 27, and 39 GeV cover a wide range of baryon chemical potential $\mu_{B}$ (100--400 MeV) in the QCD phase diagram.  Results from the BES Phase-I related to ``turn-off'' of strongly interacting quark--gluon plasma (sQGP) signatures and signals of QCD phase boundary are reported. In addition to this, an outlook is presented for the future BES Phase-II program and a possible fixed target program at STAR.
\end{abstract}
  
\section{Introduction}
At sufficiently high temperature $T$ or baryon chemical potential $\mu_{B}$ QCD predicts a phase transition from hadrons to the plasma of its fundamental constituents -- quarks and gluons~\cite{lattice1}. Search for and understanding of the nature of this transition has been a long-standing challenge to high-energy nuclear and particle physics community. In 2005, just five years after start up of RHIC, the first convincing arguments on the existence of de-confined strongly interacting partonic matter with unexpected properties of   perfect quark-gluon liquid, constituent number scaling, jet quenching and heavy-quark suppression  were revealed~\cite{qgp}.  A central goal now is to map out as much of the QCD phase diagram in $T$, $\mu_{B}$  plane as possible, trying to understand various ways in which the hadron-to-sQGP transition may occur.  STAR proposal for the BES program was published in 2010~\cite{ref_bes}. The goal is to search for the ``turn-off'' of sQGP signatures, signals of QCD phase boundary and existence of a critical point in the QCD phase diagram.  First two topics are covered in this talk, the last one is discussed  in two other STAR contributions to this conference \cite{Sahoo, Li}.  
 
The results presented here are based on the data collected by the STAR collaboration. The almost uniform acceptance for different identified particles and collision energies at midrapidity is an important advantage of the STAR detector for the BES program. The main tracking device -- the Time Projection Chamber (TPC) -- covering  2$\pi$ in azimuth ($\phi$) and $-$1 to 1 in pseudorapidity ($\eta$) provides momentum measurements as well as particle identification (PID) of charged particles.
For the higher transverse momentum ($p_{T}$) region, the Time Of Flight (TOF) detector is quite effective in distinguishing between different particle types. Particles are identified using the ionization energy loss in TPC and time-of-flight information from TOF~\cite{tpc_tof}. The centrality selection in STAR is done using the uncorrected charge particle multiplicity measured in
the TPC within $|\eta|<$ 0.5~\cite{9gev}. 

During its twelve years of operation the RHIC machine has delivered a variety of nuclear beams (Au, Cu, d). The most frequently used  c.m.s. energies per nucleon-nucleon pair were $\sqrt{s_{NN}}$ = 200  and 62.4 GeV.  The last few years have witnessed, quite naturally, a shift of experimental activity to lower energies as well as a change of the colliding  species. After few small-statistic Cu+Cu exploration runs at $\sqrt{s_{NN}}$ =22.4 GeV  in 2005 and  with Au+Au at $\sqrt{s_{NN}}$ = 9.2 GeV in 2008 \cite{9gev}, STAR collected large-statistics  data sets with Au+Au at  $\sqrt{s_{NN}}$ = 7.7, 11.5 and 39 GeV in 2010, and 19.6 and 27 GeV in 2011~\cite{lokesh}.  It is noteworthy that $\sqrt{s_{NN}}$ = 7.7 GeV, which is much below the RHIC design injection energy of 19.6 GeV, remains so far also the lowest energy achieved with hadron collider.  In 2012 RHIC collided for the first time two uranium beams  at $\sqrt{s_{NN}}$ = 193 GeV.   Left panel of  Figure~\ref{1}  shows the statistics of the recorded heavy-ion data  from all RHIC runs by the STAR detector. Significant amounts of data have been accumulated since 2010 when the fast data acquisition and the TOF subsystem upgrades were completed.

\section{``Turn-off'' of sQGP signatures}
\subsection{Supression of high-$p_{T}$ hadrons}
Important signature of sQGP at top RHIC energy is the Nuclear Modification Factor $R_{\rm{CP}}$~\cite{qgp}, which is defined as ratio of yields at central collisions to peripheral collisions, scaled by the corresponding number of binary collisions $N_{\rm{bin}}$. It has been observed that at high $p_T$, the $R_{\rm{CP}}$ of different particles is less than unity~\cite{qgp}, which is attributed to the energy loss of the partons in the dense sQGP medium. In the absence of dense medium, there may not be suppression of high $p_T$ particles. 

\begin{figure}[htbp]
\centerline{
\hspace{-5.7cm}
\includegraphics[width=.56\textwidth] {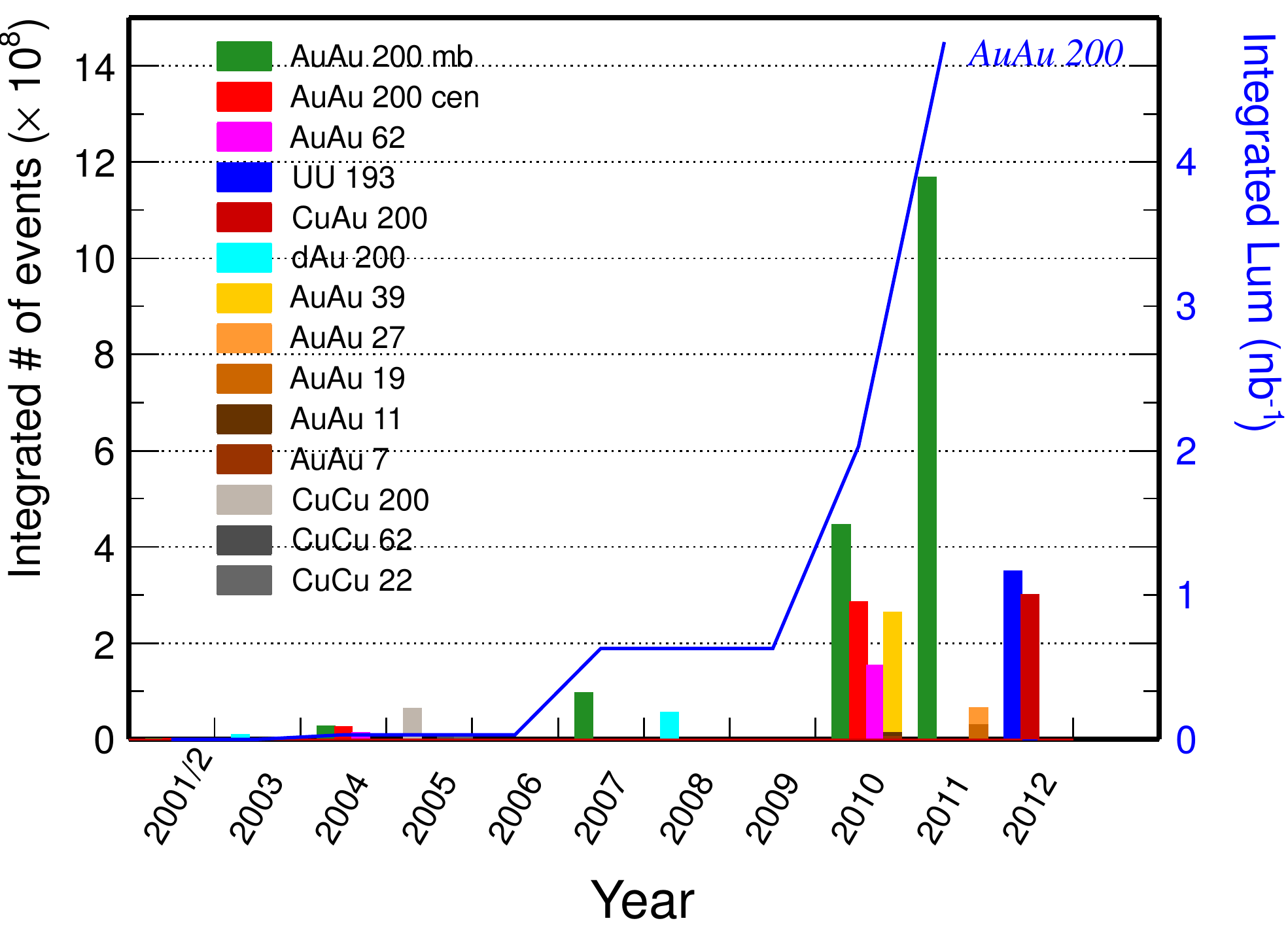}}
\vspace{-5.2cm}
\hspace{6.5cm} \includegraphics[width=.5\textwidth]{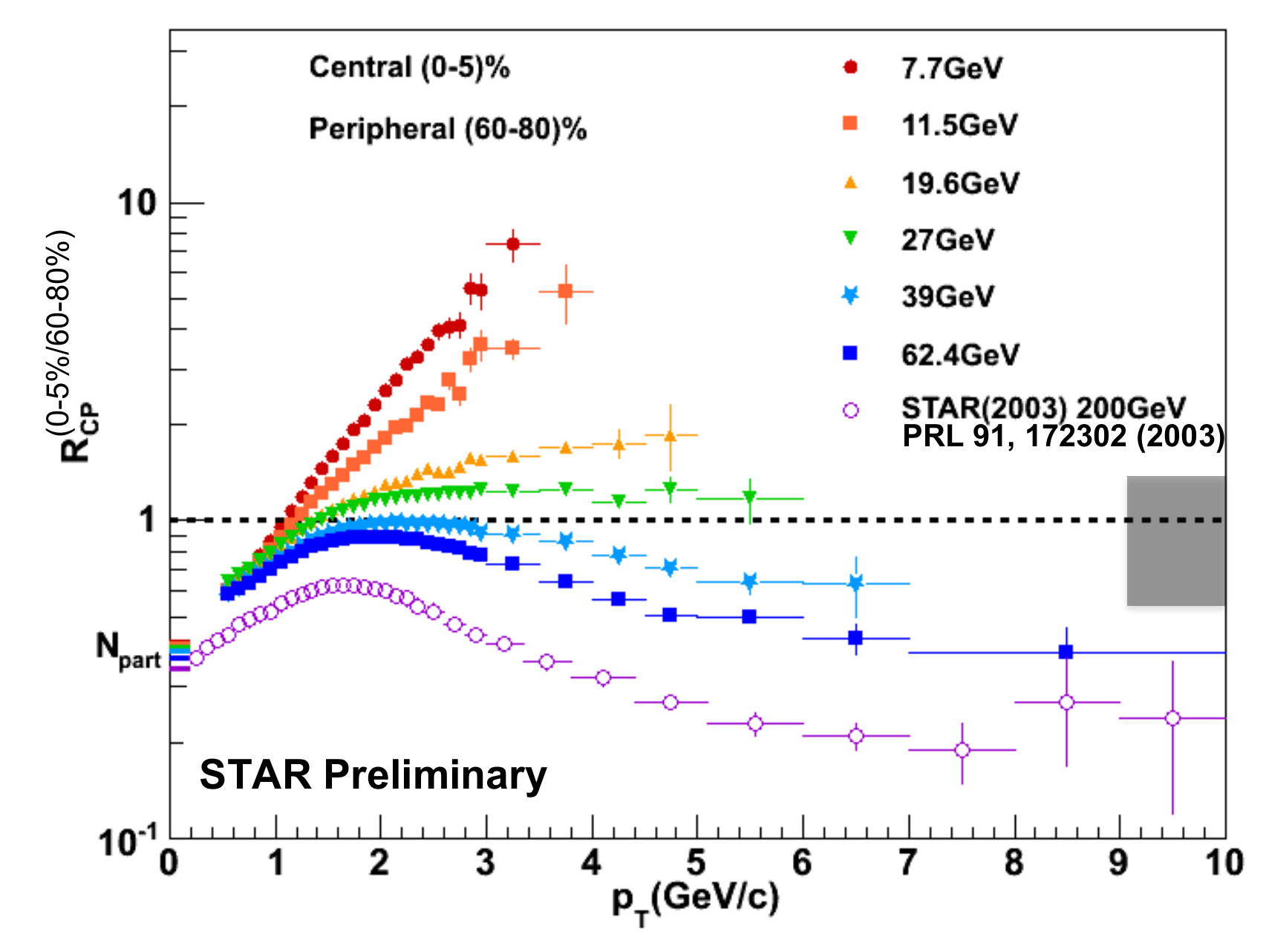}
\vspace{.0cm}
\caption[]{Left panel: Heavy-ion minimum bias/central data sets (histograms) and integrated luminosity (line) recorded by the STAR detector. Right panel: $R_{\rm{CP}}^{(0-5\%/60-80\%)}$ for charged hadrons in Au+Au collisions at
  $\sqrt{s_{NN}}=$7.7--39 GeV. Errors are statistical only. Grey band represents the normalization error from $N_{\rm{bin}}$.}
\label{1}
\vspace{-.5cm}
\end{figure}

Right panel of Figure~\ref{1} shows the $R_{\rm{CP}}$ of charged hadrons in Au+Au collisions at $\sqrt{s_{NN}}=$7.7--200 GeV. 
We observe that for $p_T>$ 2 GeV/$c$, the $R_{\rm{CP}}$  is less than unity at 39 GeV and then the value increases as the beam energy decreases. For $\sqrt{s_{NN}}>$ 27 GeV, $R_{\rm{CP}}$ is below unity, indicating a dominant role of partonic effects. Another interesting pattern is provided by the energy dependence of $R_{\rm{CP}}$ of various strange hadrons such as $K^{0}_{S}$, $\phi$, $\Lambda$, $\Xi^-$, and $\Omega^-$.  First, baryon-meson splitting observed at top RHIC energy reduces and disappears with decreasing energy. Second, for $p_T>$ 2 GeV/$c$, the $R_{\rm{CP}} (K^0_S)$  is less than unity at 39 GeV and then the value increases as the beam energy decreases. For $\sqrt{s_{NN}}<$ 19.6 GeV, $R_{\rm{CP}} (K^0_S)$ is above unity, indicating decreasing partonic effects at lower energies.  For more information on identified hadrons results see~\cite{lokesh}.

\begin{figure}[htb]
\vspace{-.5cm}
\hspace{-.2cm}
\includegraphics[width=.425\textwidth]{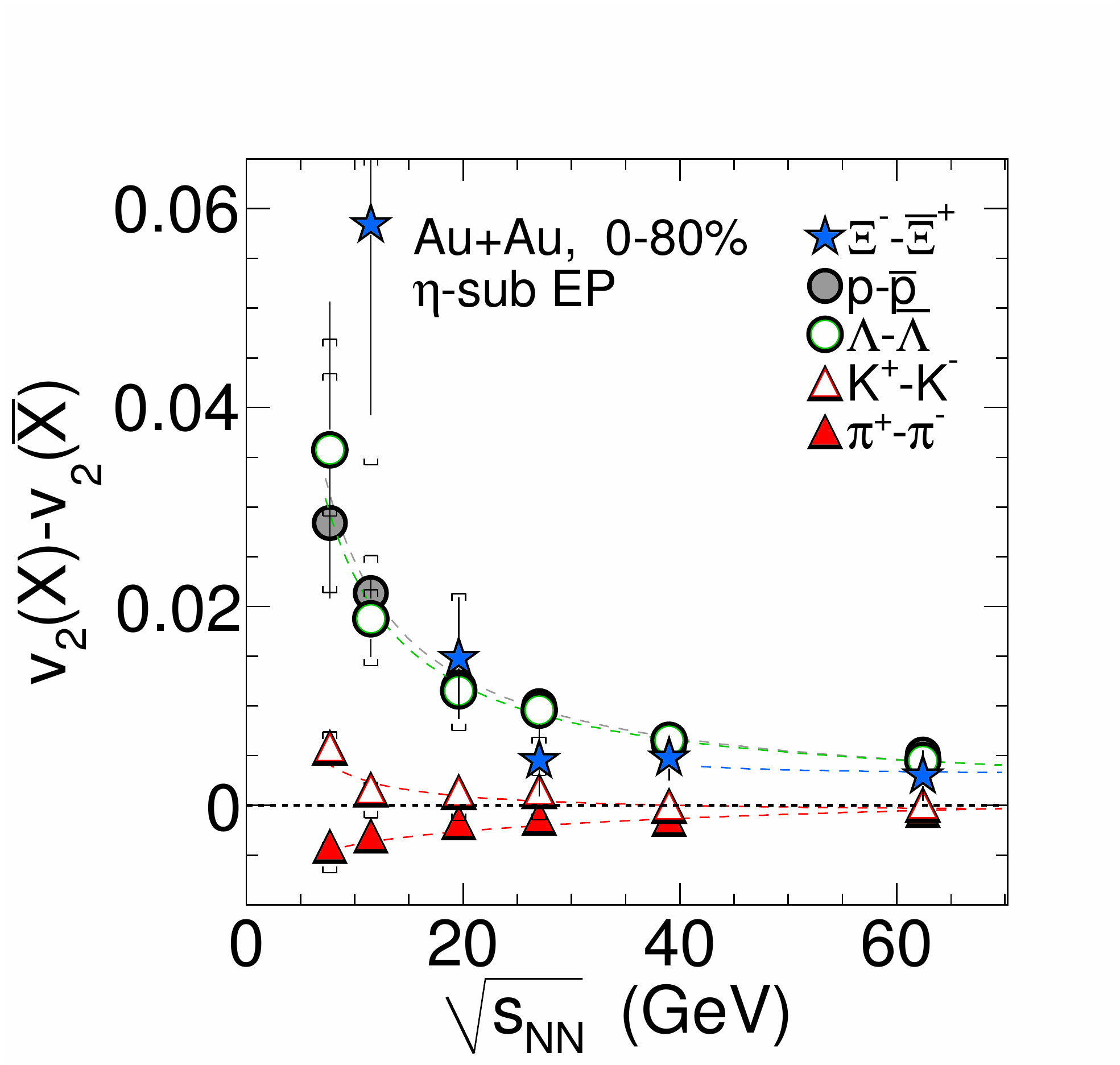}
\hspace{-.4cm}
\includegraphics[width=.675\textwidth]{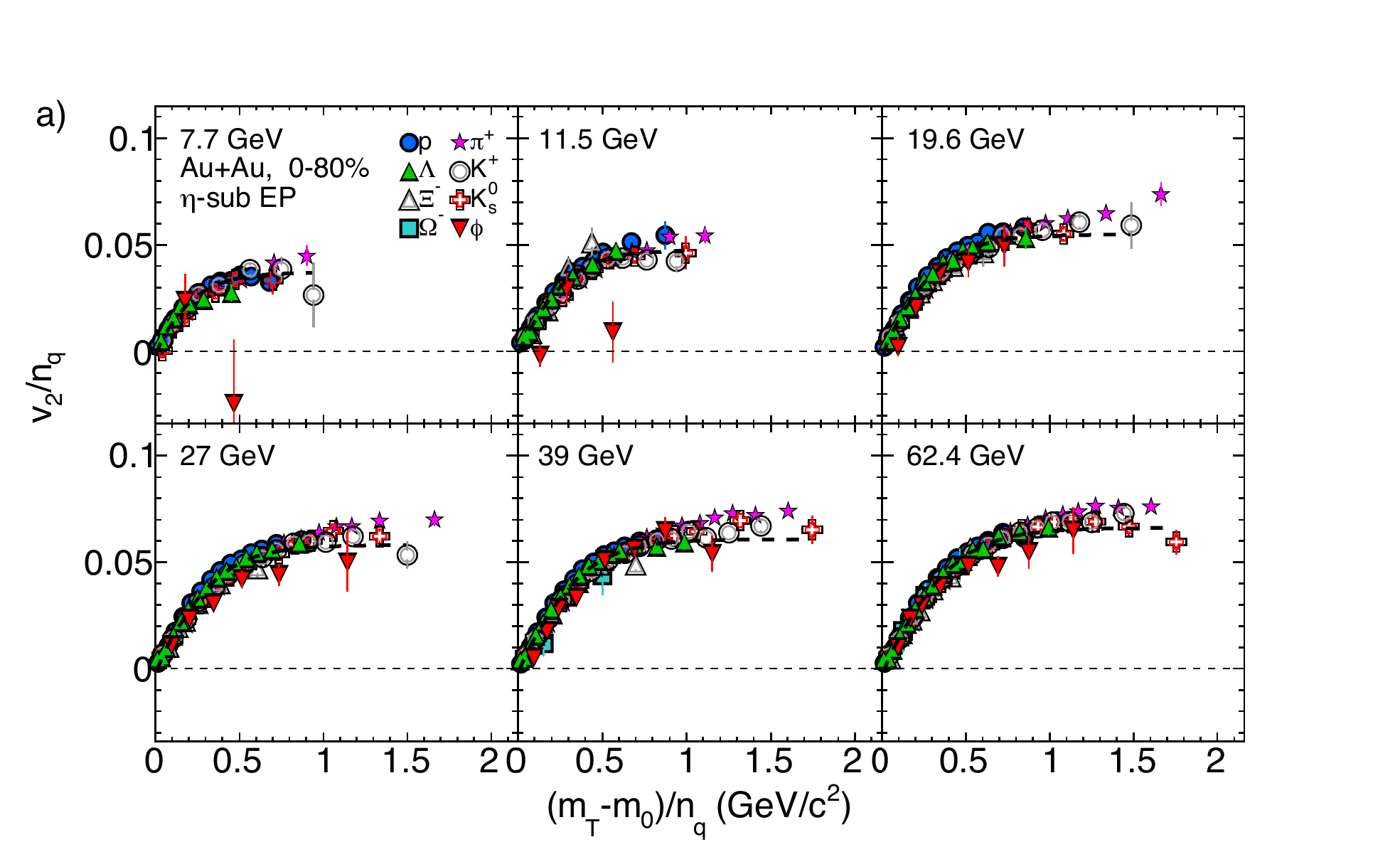}
\vspace{-.5cm}
 \caption{(Color online) Left panel: The difference in $v_{2}$ between particles ($X$) and their corresponding anti-particles (${\rm \overline{X}}$) as a function of $\sqrt{s_{NN}}$ for 0--80\% central Au+Au collisions \cite{v2_antiparticle}. The dashed lines in the plot are fits with a power-law function. 
 The error bars depict the combined statistical and systematic errors. 
 Systematic errors are plotted by caps, not combined in the error bars.
 Right panel: $v_{2}/n_{q}$ versus $(m_{T}-m_{0})/n_{q}$, for 0--80\% central Au+Au collisions for selected particles \cite{Pandit}.}
 \label{fNCQ_mT}
\vspace{-.5cm}
\end{figure}

\subsection{Elliptic Flow}
Study of the conversion of coordinate space anisotropies into momentum space anisotropies plays a central role in ongoing efforts to characterize  the transport properties of sQGP.  The  azimuthal anisotropic flow strength is usually parametrized via Fourier coefficients $v_{n} \equiv \left< \cos\left[ n (\phi - \Psi)  \right]\right>$, where $\phi$ is the azimuthal angle of the particle
and $\Psi$ is the azimuthal angle of the event plane (EP). The big surprise at RHIC came from the measurement of the $v_{2}$ coefficient, integrated elliptic flow,  which brings information on the pressure and stiffness of the equation of state during the earliest collision stages.  It was found that $v_{2}$ increases by 50\% from the  top SPS energy $\sqrt{s_{NN}}$ =17.2 GeV to the top RHIC energy $\sqrt{s_{NN}}$ =200 GeV \cite{qgp}.  The large value of $v_{2}$  observed at RHIC and later on at LHC \cite{ALICEflow} is one of the cornerstones of the perfect liquid bulk matter dynamics.  Moreover,  the differential $v_{2}(p_{T})$, that characterizes the hydrodynamic response to the initial geometry,  seems to be unchanged between the top RHIC energy and LHC energy of $\sqrt{s_{NN}}$=2.76 TeV~\cite{ALICEflow}. 
Recently published data from STAR  utilizing instead of previously used 
%event plane reconstructed in a region separated by a large pseudorapidity gap, 
$\eta$ sub--event method $v_{2}\{\rm EtaSubs\}$,  another elliptic flow method, 4--particle cumulants $v_{2}\{4\}$,  slightly modify this picture~\cite{v2_charged}.  Advantage of the cumulant method is that it removes the contribution of non--flow correlations. Systematic study of centrality, transverse momentum and pseudorapidity dependence of the inclusive charged hadron elliptic flow  $v_{2}\{4\}$ at midrapidity has shown that, as the collision energy varies from $\sqrt{s_{NN}}$ = 7.7 to 2760 GeV, the experimental data at all energies do not reveal an exact scaling. They show larger splitting in the lower $p_{T}$ region and converge at the intermediate range ($p_{T} \sim$2 GeV/$c$). The increase of $v_{2}(p_{T})$  with $\sqrt{s_{NN}}$ could be due to the change of chemical composition from low to high energies and/or larger collectivity at the higher collision energy.

STAR BES results~\cite{v2_antiparticle,Pandit}  on $v_{2}$ of identified particles in the regime where the relative contribution of 
baryon and mesons vary significantly are presented on Fig.\ref{fNCQ_mT}. In the left panel the difference in $v_{2}$ between particles 
($X$) and their corresponding anti-particles (${\rm \overline{X}}$) is plotted. A beam-energy dependent difference of the values of $v_
{2}$ between particles and corresponding anti-particles increases with decreasing beam energy and is larger for baryons compared 
to mesons. Interestingly, the flow patterns are also reflected in the number-of-constituent-quarks (NCQ) scaling of particle identified 
data. Plotting $v_{2}/n_{q}$ versus $(m_{T}-m_{0})/n_{q}$ for various particle species, where $n_{q}$ is the number of constituent 
quarks of a hadron  with mass $m_{0}$ and $m_{T}-m_{0}$ is its transverse kinetic energy, one finds the data at $\sqrt{s_{NN}}=$ 200 
GeV to collapse onto a single universal curve~\cite{qgp}.  The observed beam-energy difference between particles and anti-particles 
implies that, at lower energies, particles and anti-particles are not consistent with the universal  NCQ scaling of $v_{2}$. Such a 
breaking of the NCQ scaling could indicate increased contributions from hadronic interactions in the system evolution with 
decreasing beam energy. Nevertheless, for each given energy particles (see right panel of Fig.\ref{fNCQ_mT}) appear to follow the scaling. Note that the $v_{2}$ values for $\phi$ mesons at 7.7 and 11.5 GeV are approximately two standard deviations away from the trend defined by the other hadrons at the highest measured $p_{T}$ values.

\begin{figure}[htbp]
\begin{center}
 \includegraphics[width=.8\textwidth]{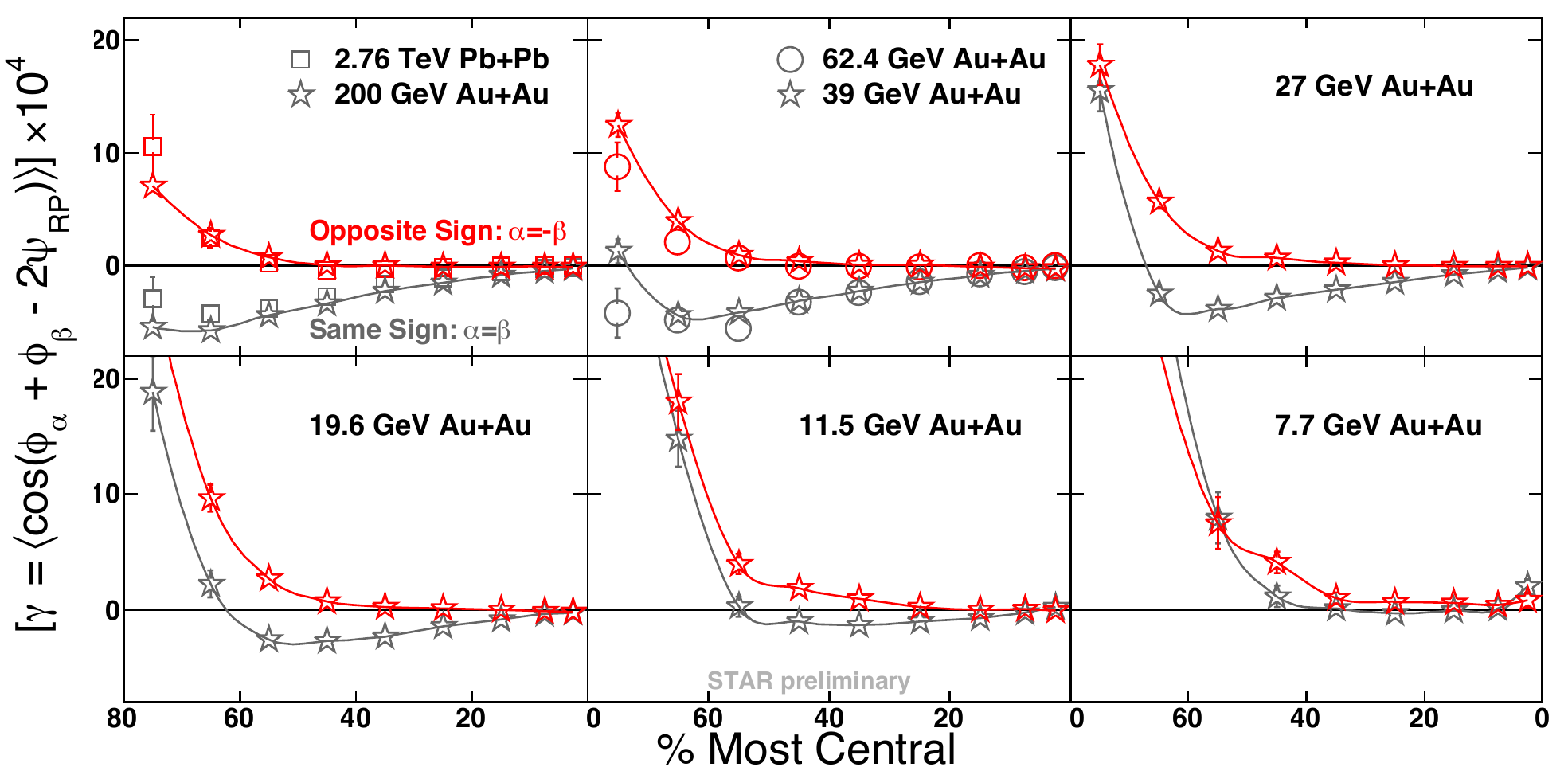}
\end{center}
\vspace{-.5cm}
\caption{(Color online) The three-point correlator, $\gamma$, as a function of centrality for Au+Au collisions 
from 200 GeV to 7.7 GeV~\cite{Dhevan}.
For comparison, we also show the results for Pb+Pb collisions at 2.76 TeV~\cite{ALICE_LPV}.
The errors are statistical only.}
\label{fig:3}
\vspace{-.5cm}
\end{figure}

\subsection{Charge separation with respect to the reaction plane }
The concept of Local Parity ($\cal P$) Violation (LPV) in high-energy heavy ion collisions
was first brought up by T.D.~Lee~\cite{Lee} and elaborated by Kharzeev {\it et al.}~\cite{Dima}.
In non-central collisions such a $\cal P$-odd domain can manifest itself via preferential
same charge particle emission for particles moving along the system's angular momentum,
due to the Chiral Magnetic Effect (CME)~\cite{Kharzeev2}.
To study this effect, a three-point mixed harmonics azimuthal correlator 
$\gamma = \langle \cos(\phi_{\alpha} + \phi_{\beta} - 2\psi_{\rm RP}) \rangle$
was proposed~\cite{Sergei2}. Here $\phi$ is the particle azimuthal angle, $\alpha$ and $\beta$ denote the particle type: $\alpha$, $\beta = +$, $-$ and $\psi_{\rm RP}$ is the reaction plane azimuth.
The observable $\gamma$ is {$\cal P$}-even, but sensitive to the fluctuation of charge separation.
STAR measurements of the correlator were reported for Au+Au and Cu+Cu collisions at 200 GeV and 62.4 GeV~\cite{STAR_LPV1,STAR_LPV2},
showing a clear difference between the opposite sign (OS) and the same sign (SS) correlations,
qualitatively consistent with the picture of CME and LPV.
Fig.~\ref{fig:3} presents the extension of the analysis to lower beam energies.
The results for Pb+Pb collisions at 2.76 TeV~\cite{ALICE_LPV} are also shown for comparison purposes.
A striking similarity exists between 2.76 TeV and 200 GeV  data.   With further decrease in $\sqrt{s_{NN}}$ a smooth transition occurs starting from the peripheral collisions.  At 7.7 GeV the difference between OS and SS signal dissappears.

\begin{figure}[htbp]
\begin{center}
 \includegraphics[width=.8\textwidth]{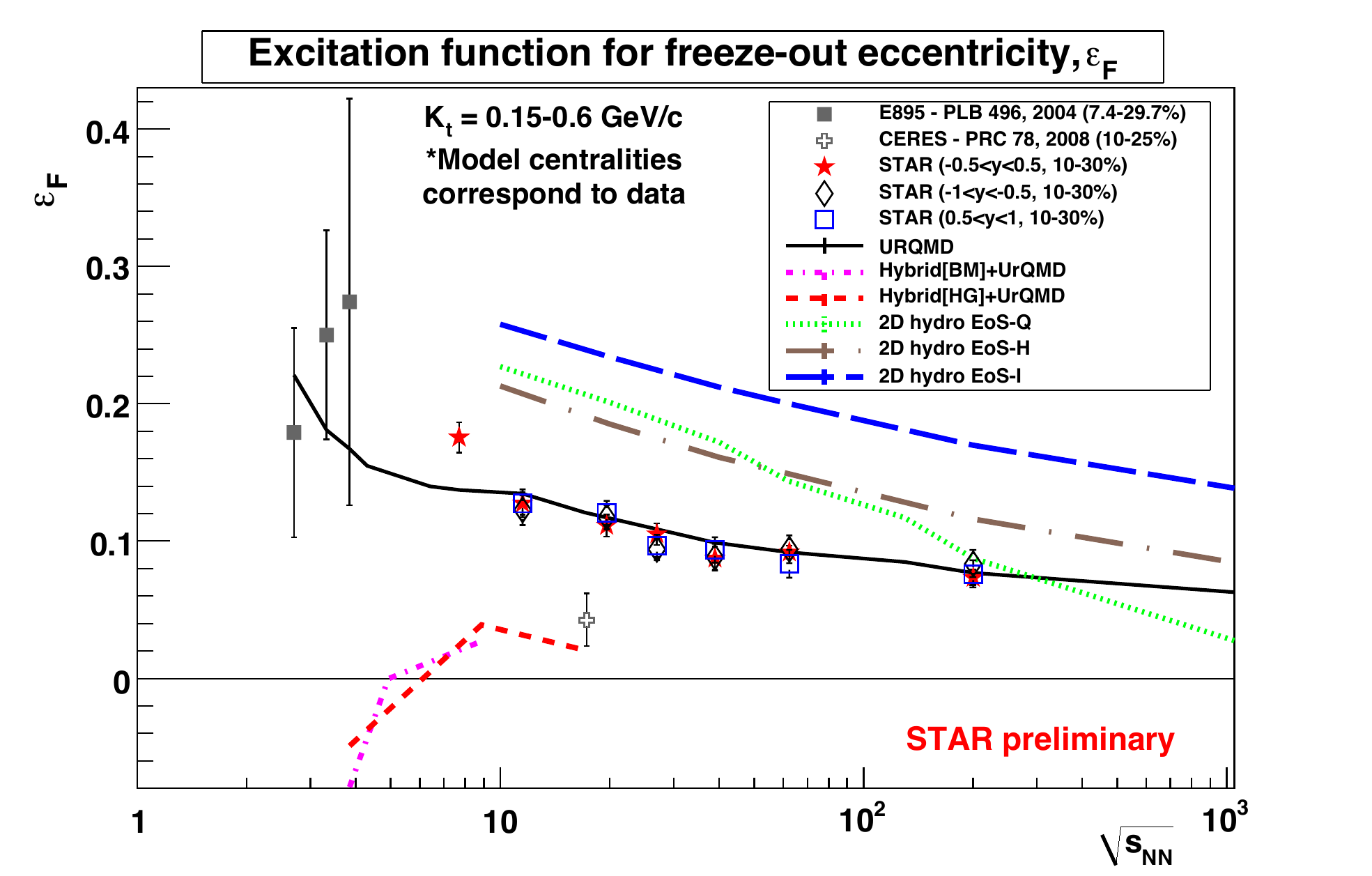}
\end{center}
\vspace{-.5cm}
\captionof{figure}{(Color online) Freeze-out eccentricity, $\varepsilon_{f}$, as a function of $\sqrt{s_{NN}}$ for data and models~\cite{Shah}.}
\label{fig:ExcFctn}
\end{figure}

\vspace{-.5cm}
\section{Signals of QCD phase boundary }
%\subsection{Azimuthal HBT}
In non-central heavy ion collisions initial eccentricity of the participant zone in the transverse plane leads to an out-of-plane extended ellipsoid. At higher energies one expects stronger pressure gradients, which would cause the shape to become more spherical. Systems with a longer lifetime, as well, would achieve a more spherical shape or could even conceivably become extended in-plane~\cite{Kolb}. Based on these two considerations, we would expect the excitation function for the freeze-out eccentricity to fall monotonically with increasing energy. The excitation function of the freeze-out eccentricity $\varepsilon_{F}$ for a centrality of 10-30$\%$ and rapidity ranges of [-0.5,0.5], [-1,-0.5] and [0.5,1] in Au+Au collisions are given in Figure~\ref{fig:ExcFctn} as a function of  
$\sqrt{s_{NN}}$. The STAR measurements are consistent with a monotonically decreasing trend. Comparison to models~\cite{AZmodel} shows that the prediction from UrQMD comes closest to describe the measurements from STAR and AGS simultaneously. The hybrid models tend to come close to the CERES point~\cite{CERES} but underpredict the rest, while 2D hydrodynamic predictions tend to overpredict the data at most energies.

\section{The BES Phase-II and STAR in Fixed Target Mode}
To strengthen the message from Phase-I, higher statistics at lower energies  is needed, especially at 7.7 and 11.5 GeV. 
Statistics for several important observables, such as $\phi$-meson $v_2$ or higher moments of net-protons distributions \cite{Li}, 
are not sufficient to draw quantitative conclusions. In order to confirm the trends between 11.5 and 19.6 GeV, 
another energy point is needed around 15 GeV in order to fill the 100 MeV gap in $\mu_B$.
This will be part of the BES Phase-II program proposed by  STAR. Simulation results indicate that with electron cooling, the luminosity could be increased by a factor of about 3--5 at 7.7 GeV and about 10 around 20 GeV~\cite{ecool}.The request for an electron cooling device at RHIC has already been submitted by STAR.  An additional improvement in luminosity may be possible by operating with longer bunches at the space-charge limit in the collider~\cite{longbunch}. Altogether  a factor or 10 improvement in luminosity is expected after these modifications. This will not only allow the precision measurements of the above important observables but will also help in the measurements of rare probes such as dileptons and hypertritons. 

To maximize the use of collisions provided by RHIC for the BES program, an option to run STAR as a fixed-target experiment
is under consideration. A fixed Au target is to be installed in the beam pipe.  This will allow the extension of the $\mu_B$ range from 400 MeV to about 800 MeV covering  thus a substantial portion of the phase diagram.  The data taking can be done concurrently during the normal RHIC running. This proposal will not affect the normal RHIC operations. Table~\ref{table} lists the proposed collision energies, corresponding fixed target center-of-mass energies, and baryon chemical potential values. The $\mu_B$ values are listed for the central collisions corresponding to fixed target \cite{Cleymans}.

  \captionof{table}{}
\vspace{-.5cm}
 \begin{center}
 \begin{tabular}{c|c|c} 
\hline
Collider mode & Fixed-target  mode & $\mu_{B}$   \\
 $\sqrt{s_{NN}}$ (GeV) &  $\sqrt{s_{NN}}$ (GeV) & (MeV) \\
\hline
19.6   & 4.5   & 585       \\% [-0.5cm]
%\hline 
15  & 4.0   &  625 \\ %[-0.5cm]
%\hline
11.5  &3.5   & 670     \\%[-0.5cm]
%\hline
7.7  &  3.0  & 720      \\%[-0.5cm]
%\hline
5  & 2.5   & 775      \\%[-0.5cm]
\hline
\end{tabular}
\end{center}
 \label{table}

\vspace{-.5cm}
\section{Summary}
Recent results from Phase-I of the RHIC BES program have substantially extended our knowledge of hot and dense de-confined QCD matter.  Interesting but smooth patterns of energy dependence are seen in most of the presented  analyses. Significant differences in particle and anti-particle $v_{2}$ coming from the high net-baryon density show up at midrapidity. This indicates increased contributions from hadronic interactions in the system evolution with decreasing beam energy.

\section*{Acknowledgements}
This work was supported by grant LA09013 of the Ministry of Education of the Czech Republic.

\end{document}